\documentclass[12pt,a4paper]{article}
\makeatletter
\def\eqnarray{%
\stepcounter{equation}%
\let\@currentlabel=\theequation
\global\@eqnswtrue
\global\@eqcnt\z@
\tabskip\@centering
\let\\=\@eqncr
$$\halign to \displaywidth\bgroup\@eqnsel\hskip\@centering
$\displaystyle\tabskip\z@{##}$&\global\@eqcnt\@ne
\hfil$\displaystyle{{}##{}}$\hfil
&\global\@eqcnt\tw@$\displaystyle\tabskip\z@{##}$\hfil
\tabskip\@centering&\llap{##}\tabskip\z@\cr}
\makeatother
\newcommand{\ket}[1]{{\vert{#1}\rangle}}

\newcommand{\fukuso}{{\mathbf C}}

\begin{document}

\title{\sl Cavity QED and Quantum Computation in the Weak Coupling Regime}
\author{
  Kazuyuki FUJII 
  \thanks{E-mail address : fujii@yokohama-cu.ac.jp },
  Kyoko HIGASHIDA 
  \thanks{E-mail address : s035577d@yokohama-cu.ac.jp },
  Ryosuke KATO 
  \thanks{E-mail address : s035559g@yokohama-cu.ac.jp }, 
  Yukako WADA 
  \thanks{E-mail address : s035588a@yokohama-cu.ac.jp }\\
  Department of Mathematical Sciences\\
  Yokohama City University\\
  Yokohama, 236--0027\\
  Japan
  }
\date{}
\maketitle
%
%
%
%
\begin{abstract}
  In this paper we consider a model of quantum computation based on n atoms of 
  laser--cooled and trapped linearly in a cavity and realize it as the n atoms 
  Tavis--Cummings Hamiltonian interacting with n external (laser) fields. 
  
  We solve the Schr{\" o}dinger equation of the model in the case of n=2 and 
  construct the controlled NOT gate by making use of a resonance condition 
  and rotating wave approximation associated to it. 
  Our method is not heuristic but completely mathematical, and the significant 
  feature is a consistent use of Rabi oscillations. 
    
  We also present an idea of the construction of three controlled NOT gates 
  in the case of n=3 which gives the controlled--controlled NOT gate.
\end{abstract}
%

%
%
%
%

\newpage

\section{Introduction}

Quantum Computation (or Computer) is a challenging task in this century 
for not only physicists but also mathematicians. 
Quantum Computation is in a usual understanding based on qubits which are 
based on two level systems (two energy levels or fundamental spins) of atoms, 
See \cite{Books} as for general theory of two level systems. 

In a realistic image of Quantum Computer we need at least one hundred atoms. 
However, then we may meet a very severe problem called Decoherence which 
destroy a superposition of quantum states in the process of unitary evolution 
of our system. At the present it is not easy to control the decoherence. 
See for example \cite{decoherence} as an introduction.  

An optical system like Cavity QED may have some advantage on this problem, 
therefore we consider a quantum computation based on Cavity QED. 
As an approximate model we realize it as the n atoms Tavis--Cummings 
Hamiltonian interacting with n external (laser) fields. As to the 
Tavis--Cummings model see \cite{Papers-1}. 
To perform the quantum computation we must first of all show that our system 
is universal \cite{nine}. 
To show it we must construct the controlled NOT operator (gate) explicitly in 
the case of $n=2$, \cite{nine}, \cite{KF1}.

For that we must embed a system of two--qubits in a space of wave functions 
of the model and solve the Schr{\" o}dinger equation. In a reduced system 
we can construct the controlled NOT by use of some resonance condition and 
the rotating wave approximation associated to it. 
Then we need to assume that the coupling constants are small enough (the weak 
coupling regime in the title). 

Next we want to construct the controlled--controlled NOT operator in the case 
of $n=3$. 
For that purpose the construction of three controlled NOT gates is required 
\footnote{In the study of Cavity QED Quantum Computation this (important) 
point is missed} 
because three atoms are trapped {\bf linearly} in the cavity, and 
we present an idea toward explicit construction. If this point will be 
overcome our system of quantum computation may become complete.

\section{A Model Based on Cavity QED}

We consider a quantum computation model based on n atoms of laser--cooled and 
trapped linearly in a cavity and realize it as the n atoms 
Tavis--Cummings Hamiltonian interacting with n external (laser) fields. 
This is of course an approximate theory. In a more realistic model we must 
add other dynamical variables such as positions of atoms and their momenta 
etc. However, since such a model is almost impossible to solve we consider 
a simple one.

Then the Hamiltonian is given by
\begin{eqnarray}
\label{eq:hamiltonian-1}
H
&=&\omega {1}_{L}\otimes a^{\dagger}a + 
\frac{\Delta}{2} \sum_{j=1}^{n}\sigma^{(3)}_{j}\otimes {\bf 1} +
g\sum_{j=1}^{n}
\left(\sigma^{(+)}_{j}\otimes a+\sigma^{(-)}_{j}\otimes a^{\dagger} \right)+
\nonumber \\
&&\sum_{j=1}^{n}h_{j}
\left(\sigma^{(+)}_{j}\mbox{e}^{i(\Omega_{j}t+\phi_{j})}+
\sigma^{(-)}_{j}\mbox{e}^{-i(\Omega_{j}t+\phi_{j})} \right)\otimes {\bf 1}
\end{eqnarray}
where $\omega$ is the frequency of radiation field, $\Delta$ the energy 
difference of two level atoms, $a$ and $a^{\dagger}$ are 
annihilation and creation operators of the field, and $g$ a coupling constant, 
$\Omega_{j}$ the frequencies of external fields which are treated as classical 
fields, $h_{j}$ coupling constants, and $L=2^{n}$. 
Here $\sigma^{(+)}_{j}$, $\sigma^{(-)}_{j}$ and $\sigma^{(3)}_{j}$ are given 
as 
\begin{equation}
\sigma^{(s)}_{j}=
1_{2}\otimes \cdots \otimes 1_{2}\otimes \sigma_{s}\otimes 1_{2}\otimes \cdots 
\otimes 1_{2}\ (j-\mbox{position})\ \in \ M(L,\fukuso)
\end{equation}
where $s$ is $+$, $-$ and $3$ respectively and 
\begin{equation}
\label{eq:sigmas}
\sigma_{+}=
\left(
  \begin{array}{cc}
    0& 1 \\
    0& 0
  \end{array}
\right), \quad 
\sigma_{-}=
\left(
  \begin{array}{cc}
    0& 0 \\
    1& 0
  \end{array}
\right), \quad 
\sigma_{3}=
\left(
  \begin{array}{cc}
    1& 0  \\
    0& -1
  \end{array}
\right), \quad 
1_{2}=
\left(
  \begin{array}{cc}
    1& 0  \\
    0& 1
  \end{array}
\right).
\end{equation}

\par \noindent
In the case of $n=2$ (which is the target through this paper) 
see the figure 1. Here we state our scenario of quantum computation. 
Each external field generates a unitary element of the corresponding qubit 
(atom) like $a\otimes b$ where $a,\ b\ \in U(2)$, while an photon inserted 
generates an entanglement among such elements like 
$\sum_{j}a_{j}\otimes b_{j}$. As a whole we obtain any element in $U(4)$.

\vspace{10mm}
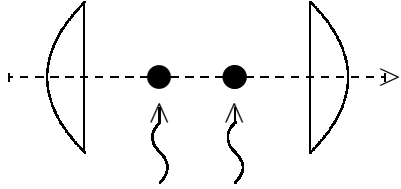
\begin{figure}
\begin{center}
\setlength{\unitlength}{1mm} 
\begin{picture}(80,40)(0,-20)
\bezier{200}(20,0)(10,10)(20,20)
\put(20,0){\line(0,1){20}}
\put(30,10){\circle*{3}}
\bezier{200}(30,-4)(32,-2)(30,0)
\bezier{200}(30,0)(28,2)(30,4)
\put(30,4){\line(0,1){2}}
\put(28.6,4){$\wedge$}
\put(40,10){\circle*{3}}
\bezier{200}(40,-4)(42,-2)(40,0)
\bezier{200}(40,0)(38,2)(40,4)
\put(40,4){\line(0,1){2}}
\put(38.6,4){$\wedge$}
\put(40,10){\circle*{3}}
\bezier{200}(50,0)(60,10)(50,20)
\put(50,0){\line(0,1){20}}
\put(10,10){\dashbox(50,0)}
\put(59,9){$>$}
\end{picture}
\vspace{-15mm}
\caption{The dotted line means a single photon inserted in the cavity and 
two curves mean external (laser) fields (which are treated as classical ones) 
subjected to atoms}
\end{center}
\end{figure}
\vspace{-10mm}

Here let us rewrite the Hamiltonian (\ref{eq:hamiltonian-1}). If we set 
\begin{equation}
\label{eq:large-s}
S_{+}=\sum_{j=1}^{n}\sigma^{(+)}_{j},\quad 
S_{-}=\sum_{j=1}^{n}\sigma^{(-)}_{j},\quad 
S_{3}=\frac{1}{2}\sum_{j=1}^{n}\sigma^{(3)}_{j},
\end{equation}
then (\ref{eq:hamiltonian-1}) can be written as 
\begin{eqnarray}
\label{eq:hamiltonian-2}
H
&=&
\omega {1}_{L}\otimes a^{\dagger}a + \Delta S_{3}\otimes {\bf 1}+ 
g\left(S_{+}\otimes a + S_{-}\otimes a^{\dagger} \right)+
\nonumber \\
&&\sum_{j=1}^{n}h_{j}
\left(\sigma^{(+)}_{j}\mbox{e}^{i(\Omega_{j}t+\phi_{j})}+
\sigma^{(-)}_{j}\mbox{e}^{-i(\Omega_{j}t+\phi_{j})} \right)\otimes {\bf 1}
\equiv H_{0}+V(t),
\end{eqnarray}
which is relatively clear. $H_{0}$ is the Tavis--Cummings Hamiltonian and we 
treat it as an unperturved one. 
We note that $\{S_{+},S_{-},S_{3}\}$ satisfy the $su(2)$--relation 
\begin{equation}
[S_{3},S_{+}]=S_{+},\quad [S_{3},S_{-}]=-S_{-},\quad [S_{+},S_{-}]=2S_{3}.
\end{equation}
However, the representation $\rho$ defined by 
\[
\rho(\sigma_{+})=S_{+},\quad \rho(\sigma_{-})=S_{-},\quad 
\rho(\sigma_{3}/2)=S_{3}
\]
is a full representation of $su(2)$, which is of course not irreducible. 

We would like to solve the Schr{\" o}dinger equation 
\begin{equation}
\label{eq:schrodinger}
i\frac{d}{dt}U=HU=\left(H_{0}+V\right)U, 
\end{equation}
where $U$ is a unitary operator. 
We can solve this equation by using the {\bf method of constant variation}. 
The equation $i\frac{d}{dt}U=H_{0}U$ is solved to be 
\[
U(t)=\left(\mbox{e}^{-it\omega S_{3}}\otimes \mbox{e}^{-it\omega N}\right)
\mbox{e}^{-itg\left(S_{+}\otimes a + S_{-}\otimes a^{\dagger}\right)}
U_{0}
\]
where $N=a^{\dagger}a$ is the number operator and $U_{0}$ a constant unitary. 
Here we have used the resonance condition 
\begin{equation}
\label{eq:old-resonance}
\omega=\Delta
\end{equation}
, see for example \cite{Papers-2}. 
By changing $U_{0}$ $\longmapsto$ $U_{0}(t)$ and substituting into 
(\ref{eq:schrodinger}) we have the equation 
\begin{equation}
\label{eq:reduction-equation}
i\frac{d}{dt}U_{0}=
\mbox{e}^{itg\left(S_{+}\otimes a + S_{-}\otimes a^{\dagger}\right)}
\left(\mbox{e}^{it\omega S_{3}}\otimes \mbox{e}^{it\omega N}\right)
V(t)
\left(\mbox{e}^{-it\omega S_{3}}\otimes \mbox{e}^{-it\omega N}\right)
\mbox{e}^{-itg\left(S_{+}\otimes a + S_{-}\otimes a^{\dagger}\right)}
U_{0}
\end{equation}
after some algebras. We would like to calculate the right hand side of 
(\ref{eq:reduction-equation}) explicitly, which is however a very hard task 
due to the term 
$\mbox{e}^{-itg\left(S_{+}\otimes a + S_{-}\otimes a^{\dagger}\right)}$. 
It has been done only for $n=1$, $2$ and $3$ as far as we know, 
\cite{Papers-3}, \cite{Papers-2}. 
The case $n=1$ which is just the Jaynes--Cummings model is not interesting 
from the point of view of a quantum computation, so we restrict to the case 
$n=2$ in the following. 

First let us write down each term (\ref{eq:reduction-equation}). 
From the result in \cite{Papers-2} and some algebras we have 
\begin{eqnarray}
\label{eq:exponential}
&&\mbox{e}^{-itg\left(S_{+}\otimes a + S_{-}\otimes a^{\dagger}\right)}=
        \nonumber \\
&&\left(
  \begin{array}{cccc}
    \frac{2N+2}{2N+3}f(N+1)+1 & -ik(N+1)a & -ik(N+1)a & 
    \frac{2}{2N+3}f(N+1)a^{2} \\
    -ik(N)a^{\dagger} & f(N)+1 & f(N) & -ik(N)a \\
    -ik(N)a^{\dagger} & f(N) & f(N)+1 & -ik(N)a \\
    \frac{2}{2N-1}f(N-1){a^{\dagger}}^{2} & -ik(N-1)a^{\dagger} & 
    -ik(N-1)a^{\dagger} & \frac{2N}{2N-1}f(N-1)+1
  \end{array}
\right)
\end{eqnarray}
where 
\[
f(N)=\frac{-1+\mbox{cos}\left(tg\sqrt{2(2N+1)}\right)}{2},\quad 
k(N)=\frac{\mbox{sin}\left(tg\sqrt{2(2N+1)}\right)}{\sqrt{2(2N+1)}},
\]
and 
\begin{eqnarray}
\label{eq:S-N-V-S-N}
&&\left(\mbox{e}^{it\omega S_{3}}\otimes \mbox{e}^{it\omega N}\right)
V(t)
\left(\mbox{e}^{-it\omega S_{3}}\otimes \mbox{e}^{-it\omega N}\right)
           \nonumber \\
=&&
\left(
  \begin{array}{cccc}
    0 & h_{2}\mbox{e}^{i\{(\Omega_{2}+\omega)t+\phi_{2}\}} & 
    h_{1}\mbox{e}^{i\{(\Omega_{1}+\omega)t+\phi_{1}\}} & 0       \\
    h_{2}\mbox{e}^{-i\{(\Omega_{2}+\omega)t+\phi_{2}\}} & 0 &
    0 & h_{1}\mbox{e}^{i\{(\Omega_{1}+\omega)t+\phi_{1}\}}       \\
    h_{1}\mbox{e}^{-i\{(\Omega_{1}+\omega)t+\phi_{1}\}} & 0 &
    0 & h_{2}\mbox{e}^{i\{(\Omega_{2}+\omega)t+\phi_{2}\}}       \\
    0 & h_{1}\mbox{e}^{-i\{(\Omega_{1}+\omega)t+\phi_{1}\}} & 
    h_{2}\mbox{e}^{-i\{(\Omega_{2}+\omega)t+\phi_{2}\}} & 0 
  \end{array}
\right)\otimes {\bf 1}. \nonumber \\
&{}& 
\end{eqnarray}
Therefore we can calculate the term 
\begin{equation}
\label{eq:full-term}
F(t)\equiv 
\mbox{e}^{itg\left(S_{+}\otimes a + S_{-}\otimes a^{\dagger}\right)}
\left(\mbox{e}^{it\omega S_{3}}\otimes \mbox{e}^{it\omega N}\right)
V(t)
\left(\mbox{e}^{-it\omega S_{3}}\otimes \mbox{e}^{-it\omega N}\right)
\mbox{e}^{-itg\left(S_{+}\otimes a + S_{-}\otimes a^{\dagger}\right)}
\end{equation}
from (\ref{eq:exponential}) and (\ref{eq:S-N-V-S-N}). However, we omit 
the explicit form because of being too complicated. 

Next let us go to a quantum computation based on two atoms of laser--cooled 
and trapped linearly in a cavity.

\section{Quantum Computation} 

Let us make a short review of two--qubits. Each element can be written as 
\[
\psi=a_{++}\ket{+}\otimes \ket{+}+a_{+-}\ket{+}\otimes \ket{-}+
a_{-+}\ket{-}\otimes \ket{+}+a_{--}\ket{-}\otimes \ket{-}
\]
with two bases $\ket{+}$ and $\ket{-}$ and $|a_{++}|^{2}+|a_{+-}|^{2}+
|a_{-+}|^{2}+|a_{--}|^{2}=1$. Here if we identify 
\[
\ket{+}=
\left(
  \begin{array}{c}
   1 \\
   0
  \end{array}
\right),\quad 
\ket{-}=
\left(
  \begin{array}{c}
   0 \\
   1
  \end{array}
\right),
\]
then $\psi$ above becomes 
\begin{equation}
\label{eq:}
\psi=
\left(
  \begin{array}{c}
    a_{++} \\
    a_{+-} \\
    a_{-+} \\
    a_{--}
  \end{array}
\right).
\end{equation}

How do we embed two--qubits in our quantized system ? It is not known at the 
moment, which will depend on some method of experimentalists. 
Therefore let us consider the simplest one like 
\begin{equation}
\label{eq:encode two qubits}
\ket{\psi(t)}=
\left(
  \begin{array}{c}
    a_{++}(t) \\
    a_{+-}(t) \\
    a_{-+}(t) \\
    a_{--}(t)
  \end{array}
\right)\otimes \ket{0}, 
\end{equation}
where $\ket{0}$ is the ground state of the radiation field ($a\ket{0}=0$). 
We note that in full theory we must consider the following superpositions 
\[
\ket{\Psi(t)}=\sum_{n=0}^{\infty}
\left(
  \begin{array}{c}
    a_{++,n}(t) \\
    a_{+-,n}(t) \\
    a_{-+,n}(t) \\
    a_{--,n}(t)
  \end{array}
\right)\otimes \ket{n} 
\]
as a wave function, which is however too complicated to solve. 

To determine a dynamics that the coefficients 
$a_{++},a_{+-},a_{-+},a_{--}$ will satisfy we substitute 
(\ref{eq:encode two qubits}) into the equation 
\begin{equation}
\label{eq:reduction-equation-application}
i\frac{d}{dt}\ket{\psi(t)}=F(t)\ket{\psi(t)}.
\end{equation}
See Appendix for the full calculations. 
The equation is not satisfied under the restrictive ansatz 
(\ref{eq:encode two qubits}). However, excited states 
$\ket{1}$, $\ket{2}$, $\ket{3}$ which have no corresponding kinetic terms
contain the coupling constants $h_{1}$ and $h_{2}$, so the equation is 
approximately satisfied if they are {\bf small enough} (namely, in the weak 
coupling regime in the title).

\par \noindent
Therefore the (full) equation is reduced to the equations of 
$\{a_{++},a_{+-},a_{-+},a_{--}\}$ at the ground state. 
\begin{eqnarray}
&&i\frac{d}{dt}a_{++}=     \nonumber \\
&&\left[
h_{1}\mbox{e}^{i\{(\Omega_{1}+\omega)t+\phi_{1}\}}
\left\{f(0)+\frac{2}{3}f(0)f(1)+k(0)k(1)\right\}+   
\right. \nonumber \\
&&\ \left.
h_{2}\mbox{e}^{i\{(\Omega_{2}+\omega)t+\phi_{2}\}}
\left\{1+f(0)+\frac{2}{3}f(1)+\frac{2}{3}f(0)f(1)+
k(0)k(1)\right\}\right]a_{+-}+    \nonumber \\
&&\left[
h_{1}\mbox{e}^{i\{(\Omega_{1}+\omega)t+\phi_{1}\}}
\left\{1+f(0)+\frac{2}{3}f(1)+\frac{2}{3}f(0)f(1)+k(0)k(1)\right\}+
\right. \nonumber \\
&&\ \left.
h_{2}\mbox{e}^{i\{(\Omega_{2}+\omega)t+\phi_{2}\}}
\left\{f(0)+\frac{2}{3}f(0)f(1)+k(0)k(1)\right\}
\right]a_{-+},
\end{eqnarray}
\begin{eqnarray}
&&i\frac{d}{dt}a_{+-}=     \nonumber \\
&&\left[
h_{1}\mbox{e}^{-i\{(\Omega_{1}+\omega)t+\phi_{1}\}}
\left\{f(0)+\frac{2}{3}f(0)f(1)+k(0)k(1)\right\}+   
\right. \nonumber \\
&&\ \left.
h_{2}\mbox{e}^{-i\{(\Omega_{2}+\omega)t+\phi_{2}\}}
\left\{1+f(0)+\frac{2}{3}f(1)+\frac{2}{3}f(0)f(1)+
k(0)k(1)\right\}\right]a_{++}+    \nonumber \\
&&\left[
h_{1}\mbox{e}^{i\{(\Omega_{1}+\omega)t+\phi_{1}\}}
\left\{1+f(0)\right\}+
h_{2}\mbox{e}^{i\{(\Omega_{2}+\omega)t+\phi_{2}\}}
f(0)\right]a_{--},
\end{eqnarray}
\begin{eqnarray}
&&i\frac{d}{dt}a_{-+}=     \nonumber \\
&&\left[
h_{1}\mbox{e}^{-i\{(\Omega_{1}+\omega)t+\phi_{1}\}}
\left\{1+f(0)+\frac{2}{3}f(1)+\frac{2}{3}f(0)f(1)+   
k(0)k(1)\right\}+
\right. \nonumber \\
&&\ \left.
h_{2}\mbox{e}^{-i\{(\Omega_{2}+\omega)t+\phi_{2}\}}
\left\{f(0)+\frac{2}{3}f(0)f(1)+k(0)k(1)\right\}
\right]a_{++}+    \nonumber \\
&&\left[
h_{1}\mbox{e}^{i\{(\Omega_{1}+\omega)t+\phi_{1}\}}
f(0)+
h_{2}\mbox{e}^{i\{(\Omega_{2}+\omega)t+\phi_{2}\}}
\left\{1+f(0)\right\}\right]a_{--},
\end{eqnarray}
\begin{eqnarray}
&&i\frac{d}{dt}a_{--}=     \nonumber \\
&&\left[
h_{1}\mbox{e}^{-i\{(\Omega_{1}+\omega)t+\phi_{1}\}}
\left\{1+f(0)\right\}+
h_{2}\mbox{e}^{-i\{(\Omega_{2}+\omega)t+\phi_{2}\}}
f(0)
\right]a_{+-}+ \nonumber \\
&&\left[
h_{1}\mbox{e}^{-i\{(\Omega_{1}+\omega)t+\phi_{1}\}}
f(0)+
h_{2}\mbox{e}^{-i\{(\Omega_{2}+\omega)t+\phi_{2}\}}
\left\{1+f(0)\right\}\right]a_{-+}
\end{eqnarray}
or in a matrix form 
\begin{equation}
\label{eq:matrix-form}
i\frac{d}{dt}
\left(
  \begin{array}{c}
    a_{++}(t) \\
    a_{+-}(t) \\
    a_{-+}(t) \\
    a_{--}(t)
  \end{array}
\right)
=
\left(
  \begin{array}{cccc}
          & \sharp & \sharp &         \\
   \sharp &        &        & \sharp  \\
   \sharp &        &        & \sharp  \\
          & \sharp & \sharp & 
  \end{array}
\right)
\left(
  \begin{array}{c}
    a_{++}(t) \\
    a_{+-}(t) \\
    a_{-+}(t) \\
    a_{--}(t)
  \end{array}
\right)
\end{equation}
where $\sharp$ is the corresponding matrix element from the above equations. 

We obtained the system of complete equations, which is still complicated. 
How do we solve it ? We use some resonance condition and the rotating wave 
approximation associated to it. Since 
\begin{eqnarray}
f(0)&=&\left\{-1+\mbox{cos}\left(tg\sqrt{2}\right)\right\}/2,\quad 
f(1)=\left\{-1+\mbox{cos}\left(tg\sqrt{6}\right)\right\}/2, 
\nonumber \\
k(0)&=&\mbox{sin}\left(tg\sqrt{2}\right)/\sqrt{2},\quad 
k(1)=\mbox{sin}\left(tg\sqrt{6}\right)/\sqrt{6},
\nonumber 
\end{eqnarray}
the products $f(0)f(1)$ and $k(0)k(1)$ contain the term 
$
\mbox{e}^{-itg(\sqrt{2}+\sqrt{6})}
$
by the Euler formulas $\mbox{cos}(\theta)=
(\mbox{e}^{i\theta}+\mbox{e}^{-i\theta})/2,\ \mbox{sin}(\theta)=
(\mbox{e}^{i\theta}-\mbox{e}^{-i\theta})/2i$. 
Noting 
\[
\mbox{e}^{i\{(\Omega_{1}+\omega)t+\phi_{1}\}}
\mbox{e}^{-itg(\sqrt{2}+\sqrt{6})}
=
\mbox{e}^{i\{(\Omega_{1}+\omega-(\sqrt{2}+\sqrt{6})g)t+\phi_{1}\}}, 
\]
we set a new resonance condition 
\begin{equation}
\label{eq:new-resonance}
\Omega_{1}+\omega-(\sqrt{2}+\sqrt{6})g=0.
\end{equation}
All terms in (\ref{eq:matrix-form}) except for the constant one 
$
\mbox{e}^{i\{(\Omega_{1}+\omega-(\sqrt{2}+\sqrt{6})g)t+\phi_{1}\}}=
\mbox{e}^{i\phi_{1}}
$ 
contain ones like $\mbox{e}^{i(t\theta+\alpha)}$ ($\theta\neq 0$), 
so we neglect all such oscillating terms (a rotating wave approximation). 
Then (\ref{eq:matrix-form}) reduces to a very simple matrix equation 
\begin{equation}
\label{eq:reduced-matrix-form}
i\frac{d}{dt}
\left(
  \begin{array}{c}
    a_{++}(t) \\
    a_{+-}(t) \\
    a_{-+}(t) \\
    a_{--}(t)
  \end{array}
\right)
=
\frac{-(\sqrt{3}-1)h_{1}}{24}
\left(
  \begin{array}{cccc}
      0   & \mbox{e}^{i\phi_{1}} & \mbox{e}^{i\phi_{1}} & 0  \\
   \mbox{e}^{-i\phi_{1}} &   0   &                      &    \\
   \mbox{e}^{-i\phi_{1}} &       &  0                   &    \\
      0   &                      &                      & 0
  \end{array}
\right)
\left(
  \begin{array}{c}
    a_{++}(t) \\
    a_{+-}(t) \\
    a_{-+}(t) \\
    a_{--}(t)
  \end{array}
\right).
\end{equation}
The solution is easily obtained to be 
\begin{eqnarray}
\label{eq:solution}
\left(
  \begin{array}{c}
    a_{++}(t) \\
    a_{+-}(t) \\
    a_{-+}(t) \\
    a_{--}(t)
  \end{array}
\right)
&=&
\mbox{exp}
\left\{
\frac{i(\sqrt{3}-1)h_{1}t}{24}
\left(
  \begin{array}{cccc}
      0   & \mbox{e}^{i\phi_{1}} & \mbox{e}^{i\phi_{1}} & 0  \\
   \mbox{e}^{-i\phi_{1}} &   0   &                      &    \\
   \mbox{e}^{-i\phi_{1}} &       &  0                   &    \\
      0   &                      &                      & 0
  \end{array}
\right)
\right\}
\left(
  \begin{array}{c}
    a_{++}(0) \\
    a_{+-}(0) \\
    a_{-+}(0) \\
    a_{--}(0)
  \end{array}
\right)           \nonumber \\
&=&
\left(
  \begin{array}{cccc}
    \mbox{cos}(\alpha t) & 
    \frac{i\mbox{e}^{i\phi_{1}}}{\sqrt{2}}\mbox{sin}(\alpha t) & 
    \frac{i\mbox{e}^{i\phi_{1}}}{\sqrt{2}}\mbox{sin}(\alpha t) & 0          \\
    \frac{i\mbox{e}^{-i\phi_{1}}}{\sqrt{2}}\mbox{sin}(\alpha t) & 
    \frac{1+\mbox{cos}(\alpha t)}{2}& \frac{-1+\mbox{cos}(\alpha t)}{2} & 0 \\ 
    \frac{i\mbox{e}^{-i\phi_{1}}}{\sqrt{2}}\mbox{sin}(\alpha t) & 
    \frac{-1+\mbox{cos}(\alpha t)}{2}& \frac{1+\mbox{cos}(\alpha t)}{2} & 0 \\
       0      &      0        &    0                  & 1
  \end{array}
\right)
\left(
  \begin{array}{c}
    a_{++}(0) \\
    a_{+-}(0) \\
    a_{-+}(0) \\
    a_{--}(0)
  \end{array}
\right)           \nonumber \\
&\equiv&
U(t)
\left(
  \begin{array}{c}
    a_{++}(0) \\
    a_{+-}(0) \\
    a_{-+}(0) \\
    a_{--}(0)
  \end{array}
\right)
\end{eqnarray}
where we have set $\alpha=\frac{\sqrt{6}-\sqrt{2}}{24}h_{1}$.
That is, we obtained the unitary operator $U(t)$. In particular, if we 
choose $t_{0}$ satisfying $\mbox{cos}(\alpha t_{0})=-1\ 
(\mbox{sin}(\alpha t_{0})=0)$, then 
\begin{equation}
U(t_{0})=
\left(
  \begin{array}{cccc}
    -1 &     &      &      \\
       &  0  &  -1  &      \\
       & -1  &   0  &      \\
       &     &      &  1
  \end{array}
\right)
=-
\left(
  \begin{array}{cccc}
     1 &     &      &      \\
       &  0  &   1  &      \\
       &  1  &   0  &      \\
       &     &      &  -1
  \end{array}
\right).
\end{equation}

At this stage we use a very skillful method \footnote{$U(t_{0})$ is 
imprimitive in the sense of \cite{BB}, so the main theorem in it says 
that our system is universal (namely, we can construct any element in $U(4)$)
. However, how to construct a unitary element explicitly is not given in 
\cite{BB}}. 
That is, we exchange two atoms (\cite{Dirac}) in the cavity 

\vspace{5mm}
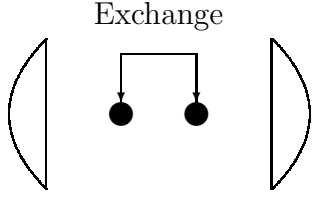
\begin{center}
\setlength{\unitlength}{1mm} 
\begin{picture}(70,40)(0,-20)
\bezier{200}(20,0)(10,10)(20,20)
\put(20,0){\line(0,1){20}}
\put(30,18){\vector(0,-1){6.5}}
\put(30,10){\circle*{3}}
\put(40,18){\vector(0,-1){6.5}}
\put(40,10){\circle*{3}}
\put(25,18){\makebox(20,10)[c]{Exchange}}
\put(30,18){\line(1,0){10}}
\bezier{200}(50,0)(60,10)(50,20)
\put(50,0){\line(0,1){20}}
\end{picture}
\end{center}
%
\vspace{-20mm}
\par \noindent
which introduces the exchange (swap) operator 
\begin{equation}
P=
\left(
  \begin{array}{cccc}
     1 &     &      &     \\
       &  0  &   1  &     \\
       &  1  &   0  &     \\
       &     &      &  1
  \end{array}
\right).
\end{equation}
Multiplying $U(t_{0})$ by $P$ gives 
\begin{equation}
PU(t_{0})=-
\left(
  \begin{array}{cccc}
     1 &     &     &     \\
       &  1  &     &     \\
       &     &  1  &     \\
       &     &     &  -1
  \end{array}
\right). 
\end{equation}
This is just the controlled $\sigma_{z}$ operator except for the overall 
constant $-1$ (an overall constant can be always neglected). 
From this it is easy to construct the controlled NOT operator, namely 
\[
C_{NOT}=({\bf 1}_{2}\otimes W)C_{\sigma_{z}}({\bf 1}_{2}\otimes W)
=
\left(
  \begin{array}{cccc}
     1 &     &     &     \\
       &  1  &     &     \\
       &     &  0  &  1  \\
       &     &  1  &  0
  \end{array}
\right)
\]
where $W$ is the Walsh--Hadamard operator given by 
\begin{equation}
W=\frac{1}{\sqrt{2}}
\left(
  \begin{array}{cc}
    1& 1 \\
    1& -1
  \end{array}
\right)
=W^{-1}.
\end{equation}
See for example \cite{KF1}. As to a construction of $W$ by making use of Rabi 
oscillations see \cite{KF2}. 

\par \noindent
Therefore our system is universal \cite{nine}, \cite{BB}. 

A comment is in order. 

\par \noindent 
(a) In the equation (\ref{eq:matrix-form}) we can set 
another resonance condition in place of (\ref{eq:new-resonance}) and 
obtain a unitary operator like $U(t)$ in (\ref{eq:solution}).

\par \noindent 
(b) In place of the ansatz (\ref{eq:encode two qubits}) we can set for example 
\[
\ket{\psi(t)}=
\left(
  \begin{array}{c}
    a_{++}(t) \\
    a_{+-}(t) \\
       0      \\
       0
  \end{array}
\right)\otimes \ket{0}
+
\left(
  \begin{array}{c}
       0      \\
       0      \\
    a_{-+}(t) \\
    a_{--}(t)
  \end{array}
\right)\otimes \ket{1}. 
\]
Then we can trace the same line shown in this section and obtain a unitary 
operator under some resonance condition like (\ref{eq:new-resonance}). 
This is a good exercise, so we leave it to the readers.

\section{Controlled-Controlled NOT Gate}

Our quantum computation model is based on n atoms of laser--cooled and 
trapped {\bf linearly} in a cavity, so we have another problem on the 
controlled NOT operators (of three types) when $n=3$. 

\par \vspace{5mm} \noindent
{\bf Problem : }Let us consider the case of three atoms in a cavity. 
How can we construct C-NOT (or C-unitary) operators for any two atoms 
among them ?  

\vspace{10mm}
\begin{figure}
\begin{center}
\setlength{\unitlength}{1mm} 
\begin{picture}(90,40)(0,-20)
\bezier{200}(20,0)(10,10)(20,20)
\put(20,0){\line(0,1){20}}
\put(30,18){\vector(0,-1){6.5}}
\put(30,10){\circle*{3}}
\put(40,18){\vector(0,-1){6.5}}
\put(40,10){\circle*{3}}
\put(50,10){\circle*{3}}
\put(25,18){\makebox(20,10)[c]{C--NOT}}
\put(30,18){\line(1,0){10}}
\bezier{200}(60,0)(70,10)(60,20)
\put(60,0){\line(0,1){20}}
\end{picture}
\end{center}
%
\vspace{-15mm}
\begin{center}
\setlength{\unitlength}{1mm} 
\begin{picture}(150,40)(0,-20)
\bezier{200}(20,0)(10,10)(20,20)
\put(20,0){\line(0,1){20}}
\put(30,18){\vector(0,-1){6.5}}
\put(30,10){\circle*{3}}
\put(40,10){\circle*{3}}
\put(50,18){\vector(0,-1){6.5}}
\put(50,10){\circle*{3}}
\put(30,18){\makebox(20,10)[c]{C--NOT}}
\put(30,18){\line(1,0){20}}
\bezier{200}(60,0)(70,10)(60,20)
\put(60,0){\line(0,1){20}}
\bezier{200}(90,0)(80,10)(90,20)
\put(90,0){\line(0,1){20}}
\put(100,10){\circle*{3}}
\put(110,18){\vector(0,-1){6.5}}
\put(110,10){\circle*{3}}
\put(120,18){\vector(0,-1){6.5}}
\put(120,10){\circle*{3}}
\put(95,18){\makebox(40,10)[c]{C--NOT}}
\put(120,18){\line(-1,0){10}}
\bezier{200}(130,0)(140,10)(130,20)
\put(130,0){\line(0,1){20}}
\end{picture}
\vspace{-15mm}
\caption{The Controlled NOT gates (of three types) for the three atoms 
in the cavity}
\end{center}
\end{figure}

\vspace{-5mm} \noindent
See the figure 2.  
These constructions are very crucial in realizing quantum logic gates, 
for example, the controlled--controlled NOT gate shown as a picture 
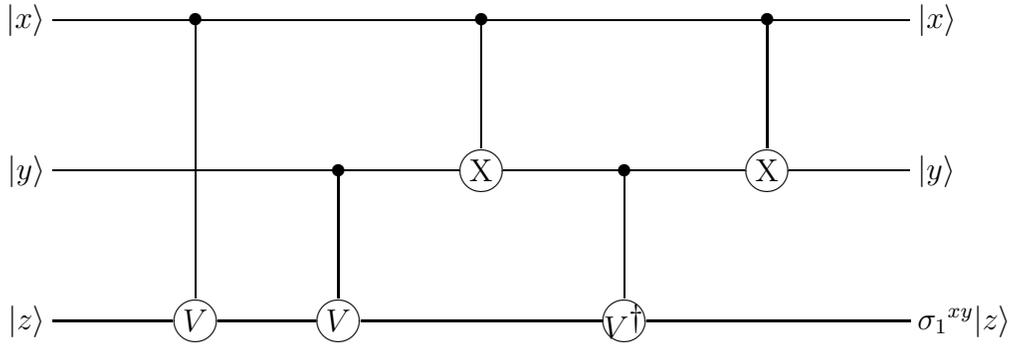
\begin{center}
\setlength{\unitlength}{1mm}  
\begin{picture}(70,50)
\put(10,40){\line(1,0){30}}   
\put(10,25){\line(1,0){30}} 
\put(10,10){\line(1,0){12}} 
\put(28,10){\line(1,0){12}}   
\put(22,35){\makebox(6,10){$\bullet$}}
\put(22,20){\makebox(6,10){$\bullet$}}
\put(25,10){\circle{6}}  
\put(22,5){\makebox(6,10){$X$}} 
\put(25,40){\line(0,-1){15}}  
\put(25,25){\line(0,-1){12}}  
\put(47,20){\makebox(20,10){$=\quad {CC}_{NOT}$}}
\end{picture}
\end{center}
\vspace{-5mm}
or in a matrix form 
\[
\left(
  \begin{array}{cccccccc}
    1 &   &    &    &    &     &    &     \\
      & 1 &    &    &    &     &    &     \\
      &   & 1  &    &    &     &    &     \\
      &   &    & 1  &    &     &    &     \\
      &   &    &    & 1  &     &    &     \\
      &   &    &    &    & 1   &    &     \\
      &   &    &    &    &     &  0 & 1   \\
      &   &    &    &    &     &  1 & 0       
  \end{array}
\right).
\]
The (usual) construction by making use of controlled NOT or controlled U gates 
is shown as a picture (\cite{KF1}, \cite{nine})
%
\begin{center}
\setlength{\unitlength}{1mm}  
\begin{picture}(140,60)
\put(10,50){\line(1,0){114}} 
\put(10,30){\line(1,0){54}}  
\put(70,30){\line(1,0){32}}  
\put(108,30){\line(1,0){16}} 
\put(10,10){\line(1,0){16}}  
\put(32,10){\line(1,0){13}}  
\put(51,10){\line(1,0){32}}  
\put(89,10){\line(1,0){35}}  
\put(0,45){\makebox(9,10)[r]{$|x\rangle$}} 
\put(0,25){\makebox(9,10)[r]{$|y\rangle$}} 
\put(0,5){\makebox(9,10)[r]{$|z\rangle$}} 
\put(125,45){\makebox(25,10)[l]{$|x\rangle$}} 
\put(125,25){\makebox(25,10)[l]{$|y\rangle$}} 
\put(125,5){\makebox(25,10)[l]{${\sigma_{1}}^{xy}|z\rangle$}} 
\put(29,13){\line(0,1){37}} 
\put(48,13){\line(0,1){17}} 
\put(67,33){\line(0,1){17}} 
\put(86,13){\line(0,1){17}} 
\put(105,33){\line(0,1){17}} 
\put(26,45){\makebox(6,10){$\bullet$}} 
\put(45,25){\makebox(6,10){$\bullet$}} 
\put(64,45){\makebox(6,10){$\bullet$}} 
\put(83,25){\makebox(6,10){$\bullet$}} 
\put(102,45){\makebox(6,10){$\bullet$}} 
\put(29,10){\circle{6}}  
\put(48,10){\circle{6}}  
\put(67,30){\circle{6}}  
\put(86,10){\circle{6}}  
\put(105,30){\circle{6}} 
\put(64,25){\makebox(6,10){X}}
\put(102,25){\makebox(6,10){X}} 
\put(26,5){\makebox(6,10){$V$}} 
\put(45,5){\makebox(6,10){$V$}} 
\put(83,5){\makebox(6,10){$V^{\mbox{\dag}}$}} 
\end{picture}
\end{center}
\vspace{-5mm}
%
where $V$ is a unitary matrix given by 
\[
V=\frac{1}{2}
\left(
  \begin{array}{cc}
    1+i& 1-i \\
    1-i& 1+i
  \end{array}
\right)
\quad \Longrightarrow \quad 
V^{2}=
\left(
  \begin{array}{cc}
    0 & 1 \\
    1 & 0 
  \end{array}
\right)
=\sigma_{1}.
\]

However, we have not seen ``realistic" constructions in any references, so 
we must give the explicit construction. 

To solve this let us state our idea. First we consider the construction of 
controlled NOT operator between the first and second atoms, namely 
\vspace{10mm}
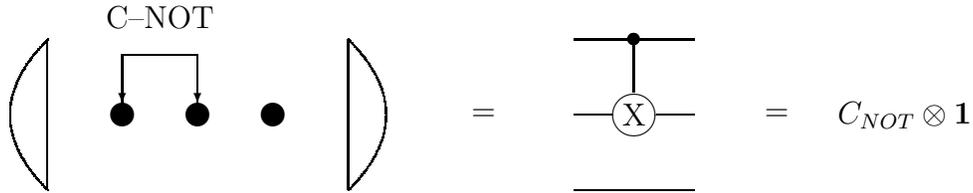
\begin{center}
\setlength{\unitlength}{1mm} 
\begin{picture}(150,40)(0,-20)
\bezier{200}(20,0)(10,10)(20,20)
\put(20,0){\line(0,1){20}}
\put(30,18){\vector(0,-1){6.5}}
\put(30,10){\circle*{3}}
\put(40,18){\vector(0,-1){6.5}}
\put(40,10){\circle*{3}}
\put(50,10){\circle*{3}}
\put(25,18){\makebox(20,10)[c]{C--NOT}}
\put(30,18){\line(1,0){10}}
\bezier{200}(60,0)(70,10)(60,20)
\put(60,0){\line(0,1){20}}
\put(90,20){\line(1,0){16}}
\put(90, 0){\line(1,0){16}}
\put(90,10){\line(1,0){5}}
\put(101,10){\line(1,0){5}}
\put(95,17){\makebox(6,6){$\bullet$}} 
\put(95,17){\makebox(6,6){$\bullet$}}
\put(98,10){\circle{6}}  
\put(95,7){\makebox(6,6){X}}
\put(98, 20){\line(0,-1){7}} 
\put(75,7){\makebox(6,6){$=$}} 
\put(114,7){\makebox(6,6){$=$}} 
\put(124,7){\makebox(20,6){$C_{NOT}\otimes {\bf 1}$}} 
\end{picture}
\end{center}
\vspace{-15mm}

Our strategy is as follows. 

\par \noindent 
(i)\ We move the third atom from the cavity. 
\par \noindent 
(ii)\ We insert a photon in the cavity as two atoms interact with it and 
subject laser fields to the atoms, and next exchange the two atoms, 
which gives the controlled NOT operator as shown in the preceding section. 
\par \noindent 
(iii)\ We return the third atom (outside the cavity) to the former position. 

\par \noindent
See the figure 3. 

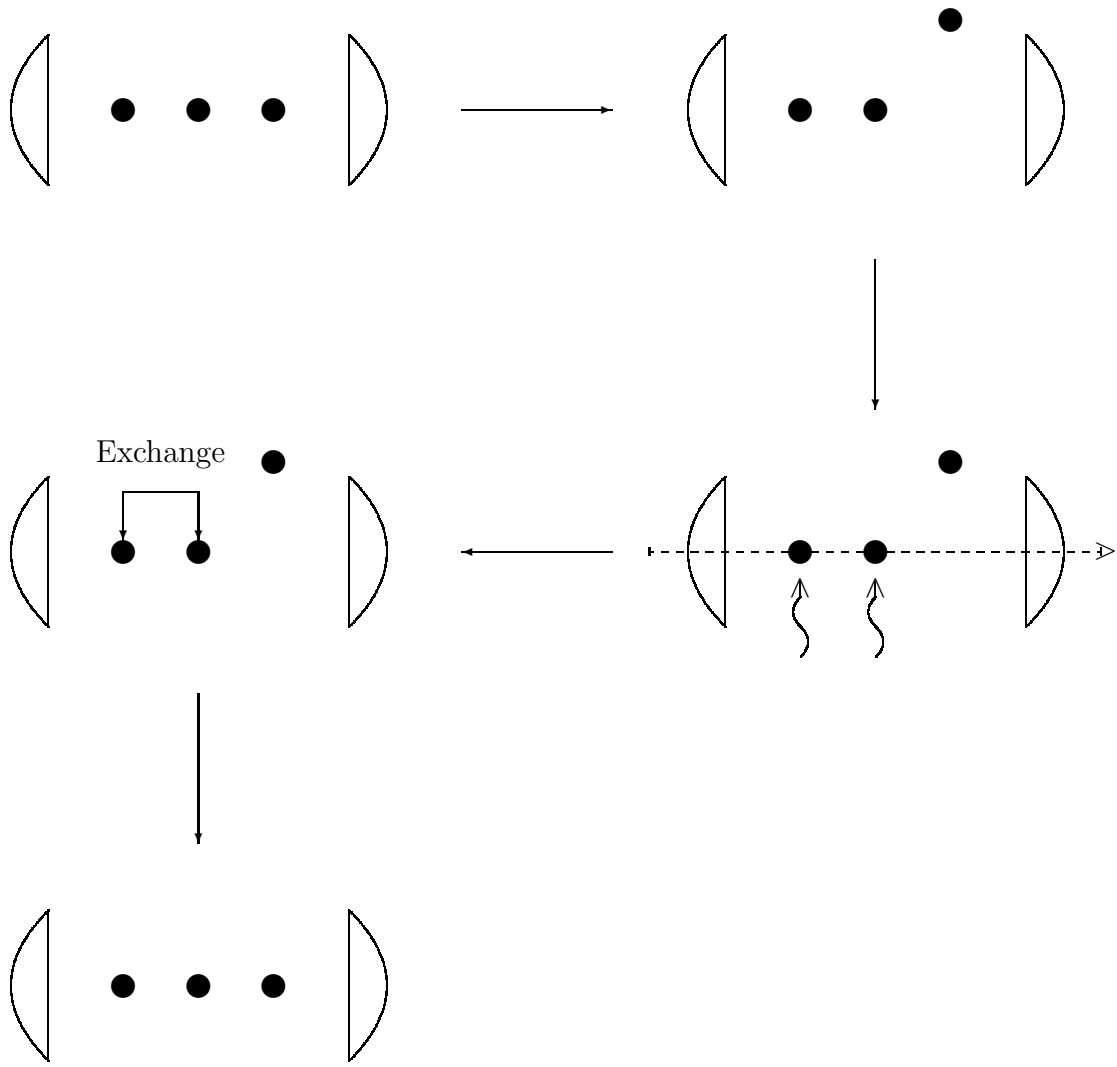
\begin{figure}
\begin{center}
\setlength{\unitlength}{1mm} 
\begin{picture}(150,40)(0,-20)
\bezier{200}(10,0)(0,10)(10,20)
\put(10,0){\line(0,1){20}}
\put(20,10){\circle*{3}}
\put(30,10){\circle*{3}}
\put(40,10){\circle*{3}}
\bezier{200}(50,0)(60,10)(50,20)
\put(50,0){\line(0,1){20}}
\put(65,10){\vector(1,0){20}}
\bezier{200}(100,0)(90,10)(100,20)
\put(100,0){\line(0,1){20}}
\put(110,10){\circle*{3}}
\put(120,10){\circle*{3}}
\put(130,22){\circle*{3}}
\bezier{200}(140,0)(150,10)(140,20)
\put(140,0){\line(0,1){20}}
\end{picture}
\end{center}
\vspace{-5mm}
\setlength{\unitlength}{1mm} 
\begin{picture}(150,20)(0,-30)
\put(125,0){\vector(0,-1){20}}
\end{picture}
\vspace{-5mm}
\begin{center}
\setlength{\unitlength}{1mm} 
\begin{picture}(150,40)(0,-20)
\bezier{200}(100,0)(90,10)(100,20)
\put(100,0){\line(0,1){20}}
\put(110,10){\circle*{3}}
\bezier{200}(110,-4)(112,-2)(110,0)
\bezier{200}(110,0)(108,2)(110,4)
\put(110,4){\line(0,1){2}}
\put(108.6,4){$\wedge$}
\put(120,10){\circle*{3}}
\bezier{200}(120,-4)(122,-2)(120,0)
\bezier{200}(120,0)(118,2)(120,4)
\put(120,4){\line(0,1){2}}
\put(118.6,4){$\wedge$}
\put(130,22){\circle*{3}}
\bezier{200}(140,0)(150,10)(140,20)
\put(140,0){\line(0,1){20}}
\put(90,10){\dashbox(60,0)}
\put(149,9){$>$}
\put(85,10){\vector(-1,0){20}}
%
%
\bezier{200}(10,0)(0,10)(10,20)
\put(10,0){\line(0,1){20}}
\put(20,18){\vector(0,-1){6.5}}
\put(20,10){\circle*{3}}
\put(30,18){\vector(0,-1){6.5}}
\put(30,10){\circle*{3}}
\put(40,22){\circle*{3}}
\put(15,18){\makebox(20,10)[c]{Exchange}}
\put(20,18){\line(1,0){10}}
\bezier{200}(50,0)(60,10)(50,20)
\put(50,0){\line(0,1){20}}
\end{picture}
\end{center}
\vspace{-5mm}
\setlength{\unitlength}{1mm} 
\begin{picture}(150,20)(0,-30)
\put(35,0){\vector(0,-1){20}}
\end{picture}
\vspace{-5mm}
\begin{center}
\setlength{\unitlength}{1mm} 
\begin{picture}(150,40)(0,-20)
\bezier{200}(10,0)(0,10)(10,20)
\put(10,0){\line(0,1){20}}
\put(20,10){\circle*{3}}
\put(30,10){\circle*{3}}
\put(40,10){\circle*{3}}
\bezier{200}(50,0)(60,10)(50,20)
\put(50,0){\line(0,1){20}}
\end{picture}
\end{center}
\vspace{-20mm}
\begin{center}
\caption{The process to construct the controlled NOT gate between the first 
atom and second one for the three atoms in the cavity}
\end{center}
\end{figure}
%

If an influence of the ``getting the third atom in and out" on the states 
space is small enough (namely, the unitary operator induced is near to the 
identity ${\bf 1}_{4}$), then we certainly obtain the controlled NOT gate 
(namely, $C_{NOT}\otimes {\bf 1}_{2}$) that we are looking for. 
Similarly we can obtain the remaining two ones. 

It is easy to generalize our idea to the $n$--atoms case. To perform a 
quantum computation we need to construct (many) controlled--controlled NOT 
gates or controlled--controlled unitary ones for three atoms among $n$--atoms, 
see \cite{nine}; $\S$7. The method is almost same, so we leave it to 
the readers. See the figure 4. 

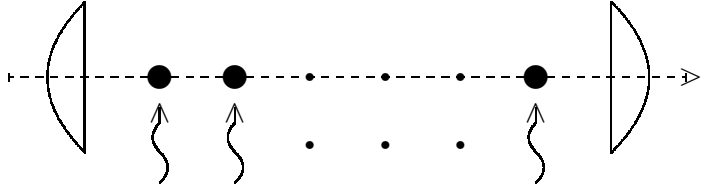
\begin{figure}
\begin{center}
\setlength{\unitlength}{1mm} 
\begin{picture}(110,40)(0,-20)
\bezier{200}(20,0)(10,10)(20,20)
\put(20,0){\line(0,1){20}}
\put(30,10){\circle*{3}}
\bezier{200}(30,-4)(32,-2)(30,0)
\bezier{200}(30,0)(28,2)(30,4)
\put(30,4){\line(0,1){2}}
\put(28.6,4){$\wedge$}
\put(40,10){\circle*{3}}
\bezier{200}(40,-4)(42,-2)(40,0)
\bezier{200}(40,0)(38,2)(40,4)
\put(40,4){\line(0,1){2}}
\put(38.6,4){$\wedge$}
\put(50,10){\circle*{1}}
\put(60,10){\circle*{1}}
\put(70,10){\circle*{1}}
\put(50,1){\circle*{1}}
\put(60,1){\circle*{1}}
\put(70,1){\circle*{1}}
\put(80,10){\circle*{3}}
\bezier{200}(80,-4)(82,-2)(80,0)
\bezier{200}(80,0)(78,2)(80,4)
\put(80,4){\line(0,1){2}}
\put(78.6,4){$\wedge$}
\bezier{200}(90,0)(100,10)(90,20)
\put(90,0){\line(0,1){20}}
\put(10,10){\dashbox(90,0)}
\put(99,9){$>$}
\end{picture}
\vspace{-15mm}
\caption{The general setting for a quantum computation based on Cavity QED. 
The dotted line means a single photon inserted in the cavity and all curves 
mean external (laser) fields (which are treated as classical ones) 
subjected to atoms}
\end{center}
\end{figure}

In principle, we can construct general quantum networks \footnote{
We must estimate an influence of the 
``getting atoms (which are not our target) in and out" on the whole states 
space, which is however difficult in our model. For that we must 
add in (\ref{eq:hamiltonian-1}) further terms necessary to calculate it}

\vspace{3mm}
By the way, a quick construction of controlled--controlled NOT gates is 
essential in general quantum networks \cite{nine}. 

We have given the exact form of evolution operator for the three atoms 
Tavis--Cummings model \cite{Papers-3}, therefore we can in principle 
track the same line shown in this paper and it may be possible to get 
the controlled--controlled NOT or many unitary gates directly (without 
combining many elementary gates like the construction of 
controlled--controlled NOT gate above). 

However, such a calculation for the three atoms case becomes very difficult 
(because we must treat $8\times 8$ matrices at each step of calculations). 
We will attempt it in the near future.

\section{Discussion}

In this paper we constructed the controlled NOT operator in the 
quantum computation based on Cavity QED which showed that our system is 
universal. We also constructed the controlled--controlled NOT operator 
(under some assumption). Therefore we can in principle perform a quantum 
computation. 

\par \noindent 
{\bf 
We expect strongly that some experimentalists will check whether our method 
works good or not. 
}

\par \noindent 
See \cite{Brune et al} and their references for some experiments on Cavity QED 
(which may be related to our method).

We conclude this paper by making a comment (which is important at least 
to us). The Tavis--Cummings model is based on (only) two energy levels of 
atoms. However, an atom has in general infinitely many energy levels, 
so it is natural to use this possibility. 
We are also studying a quantum computation based on multi--level systems of 
atoms (a qudit theory) \cite{Papers-4}. Therefore we would like to extend 
the Tavis--Cummings model based on two--levels to a model based on 
multi--levels. This is a very challenging task.

\vspace{5mm}
\noindent
{\it Acknowledgment.}
We wish to thank Shin'ichi Nojiri for his helpful comments and suggestions, 
and thank the referees for careful readings and useful suggestions. 
K. Fujii wish to thank Gilles Nogues for teaching him some experimental facts 
on Cavity QED. 

\par \vspace{10mm}
\begin{center}
 \begin{Large}
   {\bf Appendix}
 \end{Large}
\end{center}

\par \vspace{5mm} \noindent 
{\bf \ \ One Qubit Operators by Classical Fields}

Let us make a brief review of theory without the radiation field, whose 
states space is only tensor product of two level systems of each atom. 
See the figure 5. 

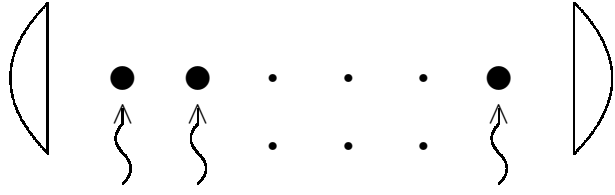
\begin{figure}
\begin{center}
\setlength{\unitlength}{1mm} 
\begin{picture}(110,40)(0,-20)
\bezier{200}(20,0)(10,10)(20,20)
\put(20,0){\line(0,1){20}}
\put(30,10){\circle*{3}}
\bezier{200}(30,-4)(32,-2)(30,0)
\bezier{200}(30,0)(28,2)(30,4)
\put(30,4){\line(0,1){2}}
\put(28.6,4){$\wedge$}
\put(40,10){\circle*{3}}
\bezier{200}(40,-4)(42,-2)(40,0)
\bezier{200}(40,0)(38,2)(40,4)
\put(40,4){\line(0,1){2}}
\put(38.6,4){$\wedge$}
\put(50,10){\circle*{1}}
\put(60,10){\circle*{1}}
\put(70,10){\circle*{1}}
\put(50,1){\circle*{1}}
\put(60,1){\circle*{1}}
\put(70,1){\circle*{1}}
\put(80,10){\circle*{3}}
\bezier{200}(80,-4)(82,-2)(80,0)
\bezier{200}(80,0)(78,2)(80,4)
\put(80,4){\line(0,1){2}}
\put(78.6,4){$\wedge$}
\bezier{200}(90,0)(100,10)(90,20)
\put(90,0){\line(0,1){20}}
%
\end{picture}
\vspace{-15mm}
\caption{The $n$ atoms in the cavity without a photon (in Figure 4)}
\end{center}
\end{figure}

The Hamiltonian in this case is 
\begin{equation}
\label{eq:hamiltonian-special}
H=
\sum_{j=1}^{n}
\left\{
\frac{\Delta}{2}\sigma^{(3)}_{j}+
h_{j}
\left(\sigma^{(+)}_{j}\mbox{e}^{i(\Omega_{j}t+\phi_{j})}+
\sigma^{(-)}_{j}\mbox{e}^{-i(\Omega_{j}t+\phi_{j})} \right)
\right\}
\end{equation}
from (\ref{eq:hamiltonian-1}), where we have omitted the unit operator 
${\bf 1}$. 
Then 
\begin{eqnarray}
H
&=&
\sum_{j=1}^{n}
\left(
\begin{array}{cc}
 \frac{\Delta}{2} & h_{j}\mbox{e}^{i(\Omega_{j}t+\phi_{j})}     \\
 h_{j}\mbox{e}^{-i(\Omega_{j}t+\phi_{j})} & -\frac{\Delta}{2} 
\end{array}
\right)_{j}       \nonumber \\
&=&
\sum_{j=1}^{n}
\left\{
\left(
\begin{array}{cc}
 \mbox{e}^{i\frac{\Omega_{j}t+\phi_{j}}{2}}  &    \\
    & \mbox{e}^{-i\frac{\Omega_{j}t+\phi_{j}}{2}}
\end{array}
\right)
\left(
\begin{array}{cc}
 \frac{\Delta}{2} & h_{j}   \\
 h_{j} & -\frac{\Delta}{2} 
\end{array}
\right)
\left(
\begin{array}{cc}
 \mbox{e}^{-i\frac{\Omega_{j}t+\phi_{j}}{2}}  &   \\
    & \mbox{e}^{i\frac{\Omega_{j}t+\phi_{j}}{2}}
\end{array}
\right)
\right\}_{j}   \nonumber \\
&=&
\left(U_{1}\otimes \cdots \otimes U_{n}\right)
\sum_{j=1}^{n}
\left(
\begin{array}{cc}
 \frac{\Delta}{2} & h_{j}   \\
 h_{j} & -\frac{\Delta}{2} 
\end{array}
\right)_{j}
\left(U_{1}\otimes \cdots \otimes U_{n}\right)^{\dagger},
\end{eqnarray}
where 
\[
U_{j}=
\left(
\begin{array}{cc}
 \mbox{e}^{i\frac{\Omega_{j}t+\phi_{j}}{2}}  &    \\
    & \mbox{e}^{-i\frac{\Omega_{j}t+\phi_{j}}{2}}
\end{array}
\right)\quad \mbox{and}\quad 
M_{j}=1_{2}\otimes \cdots \otimes 1_{2}\otimes M \otimes 1_{2}\otimes 
\cdots \otimes 1_{2}.
\]
The wave function defined by $i\frac{d}{dt}\ket{\Psi}=H\ket{\Psi}$ with 
(\ref{eq:hamiltonian-special}) can be written as a tensor product 
\begin{equation}
\ket{\Psi}=\ket{\psi_{1}}\otimes \cdots \otimes \ket{\psi_{n}},
\end{equation}
so if we define 
\[
\ket{\tilde{\Psi}}\equiv 
\left(U_{1}\otimes \cdots \otimes U_{n}\right)^{\dagger}\ket{\Psi},
\]
then it is easy to see 
\[
i\frac{d}{dt}\ket{\tilde{\Psi}}
=
\sum_{j=1}^{n}
\left(
\begin{array}{cc}
 \frac{\Delta-\Omega_{j}}{2} & h_{j}   \\
 h_{j} & -\frac{\Delta-\Omega_{j}}{2} 
\end{array}
\right)_{j}
\ket{\tilde{\Psi}}.
\]
Th solution is easy to obtain 
\[
\ket{\tilde{\Psi}(t)}=
\bigotimes_{j=1}^{n}\mbox{exp}
\left\{
-it
\left(
\begin{array}{cc}
 \frac{\Delta-\Omega_{j}}{2} & h_{j}   \\
 h_{j} & -\frac{\Delta-\Omega_{j}}{2} 
\end{array}
\right)
\right\}
\ket{\tilde{\Psi}(0)}.
\]
Therefore, the solution that we are looking for is 
\begin{eqnarray}
\ket{\Psi(t)}
&=&\left(U_{1}\otimes \cdots \otimes U_{n}\right)\ket{\tilde{\Psi}(t)}
\nonumber \\
&=&\bigotimes_{j=1}^{n}
\left(
\begin{array}{cc}
 \mbox{e}^{i\frac{\Omega_{j}t+\phi_{j}}{2}}  &    \\
    & \mbox{e}^{-i\frac{\Omega_{j}t+\phi_{j}}{2}}
\end{array}
\right)
\mbox{exp}
\left\{-it
\left(
\begin{array}{cc}
 \frac{\Delta-\Omega_{j}}{2} & h_{j}   \\
 h_{j} & -\frac{\Delta-\Omega_{j}}{2} 
\end{array}
\right)
\right\}
\ket{{\Psi}(0)}.
\end{eqnarray}

Last we note that
\[
\mbox{exp}
\left\{-it
\left(
\begin{array}{cc}
 \frac{\theta}{2} & h  \\
 h & -\frac{\theta}{2} 
\end{array}
\right)
\right\}
=
\left(
\begin{array}{cc}
  x_{11} & x_{12}  \\
  x_{21} & x_{22}
\end{array}
\right)
\]
where 
\begin{eqnarray}
x_{11}&=&\mbox{cos}\left(t\sqrt{\frac{\theta^{2}}{4}+h^{2}}\right)
-i\frac{\theta}{2}
\frac{\mbox{sin}\left(t\sqrt{\frac{\theta^{2}}{4}+h^{2}}\right)}
     {\sqrt{\frac{\theta^{2}}{4}+h^{2}}}, \nonumber \\
x_{12}&=&x_{21}=-ih
\frac{\mbox{sin}\left(t\sqrt{\frac{\theta^{2}}{4}+h^{2}}\right)}
     {\sqrt{\frac{\theta^{2}}{4}+h^{2}}}, \nonumber \\
x_{22}&=&\mbox{cos}\left(t\sqrt{\frac{\theta^{2}}{4}+h^{2}}\right)
+i\frac{\theta}{2}
\frac{\mbox{sin}\left(t\sqrt{\frac{\theta^{2}}{4}+h^{2}}\right)}
     {\sqrt{\frac{\theta^{2}}{4}+h^{2}}}. \nonumber
\end{eqnarray}

We can always construct unitary operators in $U(2)$ at each atoms by 
using Rabi osillations, see for example \cite{KF2}.

\par \vspace{10mm} \noindent 
{\bf \ \ Explicit Form of the Equation 
(\ref{eq:reduction-equation-application})}

Let us give the explicit form to (\ref{eq:reduction-equation-application}) 
for avoiding errors in the calculations. 
The left hand side of (\ref{eq:reduction-equation-application}) is 
\begin{equation}
\label{eq:left part}
\left(
  \begin{array}{c}
    i\frac{d}{dt}a_{++}(t) \\
    i\frac{d}{dt}a_{+-}(t) \\
    i\frac{d}{dt}a_{-+}(t) \\
    i\frac{d}{dt}a_{--}(t)
  \end{array}
\right)\otimes \ket{0}
\end{equation}
, while each of the right hand side becomes 
\begin{eqnarray}
\label{eq:right part-1}
&&\mbox{1-component} =h_{1}\mbox{e}^{i\{(\Omega_{1}+\omega)t+\phi_{1}\}}\times 
\nonumber \\
&&\left[
-ia_{++}\left\{k(1)+\frac{4}{5}k(1)f(2)-\frac{4}{3}f(1)k(2)\right\}\ket{1}
+a_{+-}\left\{f(0)+\frac{2}{3}f(0)f(1)+k(0)k(1)\right\}\ket{0}
\right. \nonumber \\
&&\left.
+a_{-+}\left\{1+f(0)+\frac{2}{3}f(1)+\frac{2}{3}f(0)f(1)+k(0)k(1)\right\}
\ket{0}
\right]+  \nonumber \\
&&h_{2}\mbox{e}^{i\{(\Omega_{2}+\omega)t+\phi_{2}\}}\times
\nonumber \\
&&\left[
-ia_{++}\left\{k(1)+\frac{4}{5}k(1)f(2)-\frac{4}{3}f(1)k(2)\right\}\ket{1}
\right.  \nonumber \\
&&\left.
+a_{+-}\left\{1+f(0)+\frac{2}{3}f(1)+\frac{2}{3}f(0)f(1)+k(0)k(1)\right\}
\ket{0}
\right. \nonumber \\
&&\left.
+a_{-+}\left\{f(0)+\frac{2}{3}f(0)f(1)+k(0)k(1)\right\}\ket{0}
\right], 
\end{eqnarray}
\begin{eqnarray}
\label{eq:right part-2}
&&\mbox{2-component} =h_{1}\mbox{e}^{i\{(\Omega_{1}+\omega)t+\phi_{1}\}}\times 
\nonumber \\
&&\left[
\sqrt{2}a_{++}\left\{\frac{2}{3}f(1)+k(1)k(2)+\frac{2}{3}f(1)f(2)\right\}
\ket{2}
-ia_{+-}\left\{k(0)-f(0)k(1)+k(0)f(1)\right\}\ket{1}
\right. \nonumber \\
&&\left.
-ia_{-+}\left\{k(0)-k(1)-f(0)k(1)+k(0)f(1)\right\}\ket{1}
+a_{--}\left\{1+f(0)\right\}\ket{0}
\right]+   \nonumber \\
&&h_{2}\mbox{e}^{i\{(\Omega_{2}+\omega)t+\phi_{2}\}}\times
\nonumber \\
&&\left[
\sqrt{2}a_{++}\left\{k(1)k(2)+\frac{2}{3}f(1)f(2)\right\}\ket{2}
+ia_{+-}\left\{k(1)+f(0)k(1)-k(0)f(1)\right\}\ket{1}
\right. \nonumber \\
&&\left.
+ia_{-+}\left\{f(0)k(1)-k(0)f(1)\right\}\ket{1}
+a_{--}f(0)\ket{0}
\right]+   \nonumber \\
&&
h_{1}\mbox{e}^{-i\{(\Omega_{1}+\omega)t+\phi_{1}\}}
a_{++}\left\{f(0)+\frac{2}{3}f(0)f(1)+k(0)k(1)\right\}\ket{0}+
\nonumber \\
&&
h_{2}\mbox{e}^{-i\{(\Omega_{2}+\omega)t+\phi_{2}\}}
a_{++}\left\{1+f(0)+\frac{2}{3}f(1)+\frac{2}{3}f(0)f(1)+k(0)k(1)\right\}
\ket{0},
\end{eqnarray}
\begin{eqnarray}
\label{eq:right part-3}
&&\mbox{3-component} =h_{1}\mbox{e}^{i\{(\Omega_{1}+\omega)t+\phi_{1}\}}\times 
\nonumber \\
&&\left[
\sqrt{2}a_{++}\left\{k(1)k(2)+\frac{2}{3}f(1)f(2)\right\}\ket{2}
+ia_{+-}\left\{f(0)k(1)-k(0)f(1)\right\}\ket{1}
\right. \nonumber \\
&&\left.
+ia_{-+}\left\{k(1)+f(0)k(1)-k(0)f(1)\right\}\ket{1}
+a_{--}f(0)\ket{0}
\right]+   \nonumber \\
&&h_{2}\mbox{e}^{i\{(\Omega_{2}+\omega)t+\phi_{2}\}}\times
\nonumber \\
&&\left[
\sqrt{2}a_{++}\left\{\frac{2}{3}f(1)+k(1)k(2)+\frac{2}{3}f(1)f(2)\right\}
\ket{2}
-ia_{+-}\left\{k(0)-k(1)-f(0)k(1)+k(0)f(1)\right\}\ket{1}
\right. \nonumber \\
&&\left.
-ia_{-+}\left\{k(0)-f(0)k(1)+k(0)f(1)\right\}\ket{1}
+a_{--}\left\{1+f(0)\right\}\ket{0}
\right]+   \nonumber \\
&&
h_{1}\mbox{e}^{-i\{(\Omega_{1}+\omega)t+\phi_{1}\}}
a_{++}\left\{1+f(0)+\frac{2}{3}f(1)+\frac{2}{3}f(0)f(1)+k(0)k(1)\right\}
\ket{0}+
\nonumber \\
&&
h_{2}\mbox{e}^{-i\{(\Omega_{2}+\omega)t+\phi_{2}\}}
a_{++}\left\{f(0)+\frac{2}{3}f(0)f(1)+k(0)k(1)\right\}\ket{0},
\end{eqnarray}
\begin{eqnarray}
\label{eq:right part-4}
&&\mbox{4-component} =h_{1}\mbox{e}^{i\{(\Omega_{1}+\omega)t+\phi_{1}\}}\times 
\nonumber \\
&&\left[
-2\sqrt{6}ia_{++}\left\{\frac{1}{5}k(1)f(2)-\frac{1}{3}f(1)k(2)\right\}\ket{3}
+\sqrt{2}a_{+-}\left\{\frac{2}{3}f(0)f(1)+k(0)k(1)\right\}\ket{2}
\right. \nonumber \\
&&\left.
+\sqrt{2}a_{-+}\left\{\frac{2}{3}f(1)+\frac{2}{3}f(0)f(1)+k(0)k(1)\right\}
\ket{2}
+ia_{--}k(0)\ket{1}
\right]+   \nonumber \\
&&h_{2}\mbox{e}^{i\{(\Omega_{2}+\omega)t+\phi_{2}\}}\times
\nonumber \\
&&\left[
-2\sqrt{6}ia_{++}\left\{\frac{1}{5}k(1)f(2)-\frac{1}{3}f(1)k(2)\right\}\ket{3}
+\sqrt{2}a_{+-}\left\{\frac{2}{3}f(1)+\frac{2}{3}f(0)f(1)+k(0)k(1)\right\}
\ket{2}
\right. \nonumber \\
&&\left.
+\sqrt{2}a_{-+}\left\{\frac{2}{3}f(0)f(1)+k(0)k(1)\right\}\ket{2}
+ia_{--}k(0)\ket{1}
\right]+   \nonumber \\
&&h_{1}\mbox{e}^{-i\{(\Omega_{1}+\omega)t+\phi_{1}\}}\times
\nonumber \\
&&\left[
ia_{++}\left\{k(0)-k(1)+\frac{2}{3}k(0)f(1)-2f(0)k(1)\right\}\ket{1}+
a_{+-}\left\{1+f(0)\right\}\ket{0}+a_{-+}f(0)\ket{0}
\right]+  \nonumber \\
&&h_{2}\mbox{e}^{-i\{(\Omega_{2}+\omega)t+\phi_{2}\}}\times
\nonumber \\
&&\left[
ia_{++}\left\{k(0)-k(1)+\frac{2}{3}k(0)f(1)-2f(0)k(1)\right\}\ket{1}+
a_{+-}f(0)\ket{0}+a_{-+}\left\{1+f(0)\right\}\ket{0}
\right],
\end{eqnarray}
after a long calculation by making use of (\ref{eq:full-term}).

\vspace{10mm}

\end{document}